\documentclass[twocolumn,trackchanges]{aastex63}
\usepackage{amsmath}
\usepackage{CJK}
\usepackage[]{natbib}
\usepackage[normalem]{ulem}
\usepackage{lineno}
\hypersetup{
   colorlinks,
   linkcolor={blue!88!black!80},
   citecolor={blue!88!black!80},
   urlcolor={blue!88!black!80}}

\newcommand{\kms}{km s\ensuremath{^{-1}}}

\newcommand{\msun}{\ensuremath{M_{\odot}}}

\newcommand {\hb}{\ifmmode {\rm H}\beta \else H$\beta$\fi}
\newcommand {\ha}{\ifmmode {\rm H}\alpha \else H$\alpha$\fi}

\newcommand {\mgii}{\ifmmode {\rm Mg}{\textsc{ii}} \else Mg\,{\sc ii}\fi}
\newcommand {\FeII}{Fe\,{\sc ii}}
\newcommand {\SIII}{[S\,{\sc iii}]}

\begin{document}
\begin{CJK}{UTF8}{gbsn}

\title{The Paschen Jump as a Diagnostic of the \\ Diffuse Nebular Continuum Emission in Active Galactic Nuclei\footnote{Based on observations made with the NASA/ESA Hubble Space Telescope, obtained from the Data Archive at the Space Telescope Science Institute, which is operated by the Association of Universities for Research in Astronomy, Inc., under NASA contract NAS5-26555. These observations are associated with programs 8136, 8479, 14121, 14744, 15124, and 15413.}}

\author[0000-0001-8416-7059]{Hengxiao Guo(郭恒潇)}
\affiliation{Department of Physics and Astronomy, 4129 Frederick Reines Hall, University of California, Irvine, CA, 92697-4575, USA}

\author[0000-0002-3026-0562]{Aaron J. Barth}
\affiliation{Department of Physics and Astronomy, 4129 Frederick Reines Hall, University of California, Irvine, CA, 92697-4575, USA}

\author[0000-0003-0944-1008]{Kirk T. Korista}
\affiliation{Department of Physics, Western Michigan University, 1120 Everett Tower, Kalamazoo, MI 49008-5252, USA}

\author{Michael R. Goad}
\affiliation{Department of Physics and Astronomy, University of Leicester, University Road, Leicester LE1 7RH, UK}

\author[0000-0002-8294-9281]{Edward M. Cackett}
\affiliation{Department of Physics \& Astronomy, Wayne State University, 666 W. Hancock Street, Detroit, MI 48201, USA}

\author[0000-0002-2816-5398]{Misty C.\ Bentz}
\affiliation{Department of Physics and Astronomy, Georgia State University, Atlanta, GA 30303, USA}

\author[0000-0002-0167-2453]{William N.\ Brandt}
\affiliation{Department of Astronomy and Astrophysics, 525 Davey Lab, The Pennsylvania State University, University Park, PA 16802, USA}
\affiliation{Institute for Gravitation and the Cosmos, The Pennsylvania State University, University Park, PA 16802, USA}
\affiliation{Department of Physics, 104 Davey Laboratory, The Pennsylvania State University, University Park, PA 16802, USA}

\author[0000-0002-9280-1184]{D. Gonzalez-Buitrago}
\affiliation{Universidad Nacional Aut\'onoma de M\'exico, Instituto de Astronom\'ia, AP 106,  Ensenada 22860, BC, M\'exico}

\author[0000-0003-4503-6333]{Gary J. Ferland}
\affiliation{Physics, University of Kentucky, Lexington KY 40506, USA}

\author[0000-0001-9092-8619]{Jonathan M.\ Gelbord}
\affiliation{Spectral Sciences Inc., 4 Fourth Avenue, Burlington, MA 01803}

\author[0000-0001-6947-5846]{Luis C. Ho}
\affiliation{Kavli Institute for Astronomy and Astrophysics, Peking University, Beijing 100871, China}
\affiliation{Department of Astronomy, School of Physics, Peking University, Beijing 100871, China}

\author[0000-0003-1728-0304]{Keith Horne}
\affiliation{SUPA School of Physics and Astronomy, University of St~Andrews, North Haugh, St~Andrews, KY16~9SS, Scotland UK}

\author[0000-0003-0634-8449]{Michael D. Joner}
\affiliation{Department of Physics and Astronomy, N283 ESC, Brigham Young University, Provo, UT 84602, USA}

\author[0000-0002-2180-8266]{Gerard A.\ Kriss}
\affiliation{Space Telescope Science Institute, 3700 San Martin Drive, Baltimore, MD 21218, USA}

\author[0000-0002-0151-2732]{Ian McHardy}
\affiliation{Department of Physics and Astronomy, University of Southampton, SO17 1BJ, UK}

\author[0000-0002-4992-4664]{Missagh Mehdipour}
\affiliation{Space Telescope Science Institute, 3700 San Martin Drive, Baltimore, MD 21218, USA}

\author[0000-0001-9877-1732]{Daeseong Park}
\affiliation{Department of Astronomy and Atmospheric Sciences, Kyungpook National University, Daegu, 41566, Republic of Korea}

\author[0000-0002-0164-8795]{Raymond Remigio}
\affiliation{Department of Physics and Astronomy, 4129 Frederick Reines Hall, University of California, Irvine, CA, 92697-4575, USA}

\author[0000-0002-1912-0024]{Vivian U}
\affiliation{Department of Physics and Astronomy, 4129 Frederick Reines Hall, University of California, Irvine, CA, 92697-4575, USA}

\author[0000-0001-9191-9837]{Marianne Vestergaard}
\affiliation{DARK, Niels Bohr Institute, University of Copenhagen, Jagtvej 155, DK-2200 Copenhagen N}
\affiliation{Steward Observatory and Department of Astronomy, University of Arizona, 933 N. Cherry Avenue, Tucson AZ 85721}

\begin{abstract}

Photoionization modeling of active galactic nuclei (AGN) predicts that diffuse continuum (DC) emission from the broad-line region makes a substantial contribution to the total continuum emission from ultraviolet through near-infrared wavelengths. Evidence for this DC component is present in the strong Balmer jump feature in AGN spectra, and possibly from reverberation measurements that find longer lags than expected from disk emission alone. However, the Balmer jump region contains numerous blended emission features, making it difficult to isolate the DC emission strength. In contrast, the Paschen jump region near 8200~\AA\ is relatively uncontaminated by other strong emission features. Here, we examine whether the Paschen jump can aid in constraining the DC contribution, using Hubble Space Telescope STIS spectra of six nearby Seyfert 1 nuclei. The spectra appear smooth across the Paschen edge, and we find no evidence of a Paschen spectral break or jump in total flux. We fit multi-component spectral models over the range $6800-9700$~\AA\ and find that the spectra can still be compatible with a significant DC contribution if the DC Paschen jump is offset by an opposite spectral break resulting from blended high-order Paschen emission lines. The fits imply DC contributions ranging from $\sim10\%$ to 50\% at 8000~\AA, but the fitting results are highly dependent on assumptions made about other model components. These degeneracies can potentially be alleviated by carrying out fits over a broader wavelength range, provided that models can accurately represent the disk continuum shape, \ion{Fe}{2} emission, high-order Balmer line emission, and other components.

\end{abstract}

\section{Introduction} 
The standard model for active galactic nuclei (AGN) posits the existence of a geometrically thin, optically thick accretion disk \citep{Shakura73} around a supermassive black hole (SMBH) as the primary source of continuum emission across ultraviolet (UV) through optical wavelengths. However, the application of the standard thin disk model to normal AGNs is still under debate, and many aspects of AGN accretion physics and emission processes remain poorly understood \citep[e.g.,][]{Collin02,Kishimoto08,Antonucci13,Antonucci15,Lawrence18}.

Although enormous progress has been made recently in interferometric observations that can resolve the sub-parsec infrared emission-line regions of quasars \citep{GravityCollaboration18}, the primary UV-optical emitting regions of AGN accretion disks have angular sizes too small to be resolved by any current facility. Fortunately, two indirect approaches, i.e., microlensing of gravitationally lensed quasars \citep[e.g.,][]{Morgan10} and continuum reverberation mapping \citep[see a recent review;][]{Cackett21}, are capable of resolving structure on the scale of the accretion disk by making use of time-domain information. The former utilizes multi-band photometry of microlensed quasars to constrain the temperature profiles of accretion disks based on the microlensing-perturbation-induced variations of multi-band flux ratios against wavelength \citep[e.g,][]{Poindexter08,Morgan10,Blackburne11}. The latter measures the inter-band time lags by conducting photometric monitoring campaigns to infer the disk size within the context of the ``lamp-post reprocessing'' model \citep[e.g.,][]{Cackett07}, in which UV-optical variability is interpreted as resulting from disk reprocessing of X-ray emission from a corona located above the disk's central regions.  Intensive disk reverberation mapping (IDRM) programs have successfully resolved UV-optical continuum reverberation lags in a growing number of sources in recent years though the connection of X-ray to UV/optical is unclear \citep[e.g.,][]{Mchardy14,Shappee14,Edelson15,Fausnaugh16,Edelson17,Kokubo18,Fausnaugh18,Edelson19,Cackett18,Cackett20,HernandezSantisteban20,Lobban20}.

A key result of those continuum reverberation mapping (RM) campaigns is that the UV-optical-NIR lags generally follow a trend consistent with the expected $\tau \propto \lambda^{4/3}$ dependence for reprocessing by a thin disk, but typically with a normalization $\sim$ 3 times larger than the prediction from the standard disk model, suggesting the possibility that disks may be substantially larger than expected. Similarly, disk sizes measured from microlensed quasars also indicate larger radii than anticipated from standard disk models \citep[e.g.,][]{Morgan10,Blackburne11,Mosquera13}. Suggested explanations for the long continuum reverberation lags have included modifications to accretion disk structure and reprocessing geometry \citep{Gardner17} or models in which UV-optical variations originate from temperature fluctuations within the disk itself rather than from reprocessing \citep[e.g.,][]{Cai18,Sun20a,Sun20b}, although \citet{Kammoun21} have argued that lamp-post reprocessing by a standard Novikov-Thorne disk model \citep{Novikov73} can reproduce the observed lag spectra with a more extended corona height than usually assumed.

Another possibility is that continuum emission originating from spatial scales larger than the accretion disk may be responsible for reverberation lags in excess of simple model predictions.
It is well known that the UV-optical continua of AGN include reprocessed emission from the broad-line region (BLR) in addition to the dominant disk emission \citep{Malkan82,Wills85,Maoz93,Korista01}. The most visible evidence of this BLR emission is the contribution of hydrogen free-bound emission to the ``small blue bump'' feature spanning $\sim2200-4000$~\AA, which also includes \ion{Fe}{2} emission blends and other features. The strong Balmer continuum emission below the Balmer jump at 3647~\AA\ (all wavelengths are in vacuum through this paper) is just one portion of the overall nebular diffuse continuum (DC) emission, which consists of free-free, free-bound, and scattered continuum emission spanning all wavelengths from the UV through near-infrared \citep{Korista01}. Recently, the  DC emission has been examined in detail by \citet{Lawther18} and \citet{Korista19}, who carried out photoionization modeling to assess the flux spectrum and lag spectrum of DC emission over a broad range of physical conditions of BLR clouds, finding that the DC contribution to the total continuum emission can be as high as $\sim40\%$ at wavelengths below the Balmer jump. Since the DC emission arises from the BLR, its reverberation response will introduce an additional delay signal to measured continuum lags. Continuum reverberation mapping campaigns have identified a distinct excess in lag in the $U$-band spectral region in several objects that has been identified with this DC emission \citep[e.g.,][]{Edelson15, Fausnaugh16, Cackett18, Cackett20, HernandezSantisteban20}. If the observed $U$-band excess lags do arise from DC emission from the BLR, this implies that the continuum lags across all wavelengths are also affected by the DC component, and other AGN monitoring results have suggested a substantial or even dominant contribution of DC emission to AGN optical variability \citep{Chelouche19,Cackett21b,Netzer21}.

Even if the DC component does not dominate the optical continuum reverberation lags, it certainly contributes to the lag spectrum, and determining the magnitude of its contribution is essential for isolating the wavelength-dependent lags of the accretion disk itself. To model precisely the impact of the DC component on the lag spectrum, we need to isolate the contributions of the DC and disk emission in total flux as well as in the wavelength-dependent variability. In the optical, the strongest feature in the DC spectrum appears at the Balmer jump. However, the Balmer jump spectral region contains a multitude of other emission lines including \ion{Fe}{2} emission blends as well as high-order Balmer emission lines \citep{Wills85}, making it difficult to obtain a unique determination of the DC emission strength in total flux. 

The Paschen jump at 8206 \AA\ provides another possible diagnostic of the DC contribution that is easily accessible to observations for low-redshift AGN. In contrast to the complex blend of emission features in the small blue bump region, the Paschen jump is relatively uncontaminated by other emission lines. The strength of the Paschen jump in DC spectra is expected to be much weaker than that of the Balmer jump \citep{Lawther18, Korista19}, and it has consequently received much less attention than the Balmer jump region as a diagnostic of nebular continuum emission, but with data of high S/N the Paschen spectral region could still allow for a useful and independent assessment of the DC strength that avoids the degeneracies inherent in modeling the small blue bump region.  \citet{Malkan82} proposed that the Paschen jump strength in AGN should be ``almost completely washed out'' as a result of dilution by the featureless continuum and the blended flux from broadened high-order Paschen emission lines. Higher-quality data from later near-infrared spectroscopic programs \citep[e.g.,][]{Osterbrock92, Rodriguez-Ardila00} also indicated that there was no obvious Paschen jump in Seyfert 1 galaxy spectra, although these programs did not derive quantitative constraints on the nebular continuum contribution. 

Prompted by recent developments in AGN continuum reverberation mapping, we have revisited the question of whether the DC Paschen jump is detectable in AGN spectra. In this work, we use spectra of nearby AGN from the Space Telescope Imaging Spectrograph (STIS) on the Hubble Space Telescope (HST) to fit multi-component models to the Paschen jump spectral region and examine the constraints that can be derived on the DC contribution. While there are only a small number of existing STIS spectra covering the Paschen jump in unobscured Seyfert 1, these spectra have the advantage (compared with ground-based spectra) of excluding nearly all host galaxy starlight, thanks to the 0\farcs1 or 0\farcs2 slit width of STIS, and this largely eliminates degeneracies between AGN and stellar continuum components when fitting models to the data. We describe our sample selection and spectral decomposition method in \S \ref{sec:sample}. The fitting results and DC contributions are demonstrated in \S \ref{sec:results}. Finally, we discuss the implications of the spectral modeling in \S \ref{sec:dis} and conclude in \S \ref{sec:con}.


\section{Sample and Spectral Decomposition} \label{sec:sample}
\subsection{Sample, observations, and reductions}

The sample for this work is based on the subset of reverberation-mapped AGN \citep{BentzKatz15} at redshifts $z$ $<$ 0.1  to ensure that continuum redward of the Paschen jump and the Pa$\epsilon$ line (important for constraining the contribution of high-order Paschen series) falls within the STIS G750L band. A search of the HST archive for STIS observations with the G750L grating yielded a sample of six nearby Seyfert 1 galaxies (see Table \ref{tab:sample}). In addition to these six objects, there are also archival G750L spectra of a few reverberation-mapped radio-loud sources (including 3C 120, 3C 382, and 3C 390.3), but we did not include these in our sample in order to avoid objects in which jet emission might contribute to the optical continuum.


Most of our targets were observed by HST in a single visit except for Mrk 110 and NGC 4593 (see Table \ref{tab:sample}). In order to measure the wavelength-dependent lags and constrain the DC contribution, Mrk 110 and NGC 4593 were  observed over multiple visits coordinated with recent reverberation mapping campaigns. Mrk 110 \citep{Vincentelli21} was observed on three occasions (2017 Dec 25, 2018 Jan 3, and 2018 Jan 10), while NGC 4593 \citep{Cackett18} was observed approximately daily from 2016 Jul 12 to 2016 Aug 6 with 26 successful observations. 

Spectra obtained with the STIS G750L grating cover the wavelength range from 5240 to 10270~\AA\ with an average dispersion of 4.92~\AA/pixel, and a spatial scale of 0\farcs05 pixel$^{-1}$. We downloaded the raw STIS data and calibration files from the HST archive and reprocessed the data to improve cosmic-ray and bad pixel cleaning and to apply fringing corrections. We applied charge transfer inefficiency corrections to the data with the stis\_cti  package\footnote{\url{https://www.stsci.edu/hst/instrumentation/stis/data-analysis-and-software-tools/pixel-based-cti}} in addition to the standard pipeline. STIS CCD spectra taken at wavelengths $>$7000~\AA\ are impacted by fringing, caused by interference of multiple reflections between two surfaces of the CCD \citep{Goudfrooij98}. We defringed the G750L spectra with contemporaneously obtained fringe flats according to standard STIS data reduction procedures.\footnote{\url{https://stistools.readthedocs.io/en/latest/defringe_guide.html}}

One-dimensional spectra were extracted from the wavelength- and flux-calibrated CCD frames using an extraction width of 0.35 arcsec. For each AGN other than NGC 4593 and NGC 3227, we combined the individual extracted spectra to obtain a mean flux spectrum and error spectrum to be used for spectral fitting. For NGC 4593, we found that the quality of the fringe correction varied over the monitoring campaign, and we chose to use only the epoch having the cleanest fringe correction (2016 Jul 21), rather than averaging together data from different visits.

The calibrated spectra were then corrected for Galactic extinction (see Table \ref{tab:sample}) based on the \citet{Schlegel98} dust map and the \citet{Fitzpatrick99} extinction law assuming $R_{V}=3.1$, and transformed to the AGN rest frame.

All of these AGN also have exposures in STIS UV and optical settings obtained contemporaneously with the G750L data. For this work, we primarily make use of the G750L spectra. While fits to a broader spectral range can further constrain the DC contribution, broad-band spectral models are subject to a variety of other degeneracies and ambiguities related to the spectral shape of the accretion disk continuum, the emission from \ion{Fe}{2} blends, and numerous other emission features that would need to be modeled in order to obtain precise fits to the small blue bump spectral region. Our focus in this paper is on the more restricted question of whether the DC component can be detected and constrained in the Paschen jump region, and we defer the larger and more challenging problem of broad-band spectral modeling to future work.

\begin{deluxetable*}{lccccccccc}[htb]
\tablecaption{Sample information \label{tab:sample}}
\tablecolumns{6}
\tablewidth{0pt}
\tablehead{
\colhead{Name} &
\colhead{Obs. date} &
\colhead{Exp. time $\times$ n } &
\colhead{Slit width} &
\colhead{PID} &
\colhead{$z$} &
\colhead{log $M_{\rm BH}$ }&
\colhead{log $L_{\rm 5100,AGN}$}&
\colhead{$\dot{m}$}&
\colhead{$A_{\rm V}$}\\
\colhead{} &
\colhead{} &
\colhead{(s)} &
\colhead{(\arcsec)} &
\colhead{} &
\colhead{}&
\colhead{(\msun)}&
\colhead{ (erg s$^{-1}$)}&
\colhead{}&
\colhead{(mag)}
}
\startdata
Mrk 110 &	2017 Dec 25 -- 2018 Jan 10	&60  $\times$ 9   & 0.2 & 15413 &0.035&7.29 &43.62&0.157 &0.021	\\
Mrk 493 &	2017 Aug 28	                &532 $\times$ 7   & 0.2 & 14744 &0.031&6.12 &43.11&0.718 &0.065 \\
Mrk 509 &	2017 Oct 22	                &50  $\times$ 4   & 0.2 & 15124 &0.034&8.05 &44.13&0.009 &0.152 \\   
NGC 3227&   2000 Feb 8                    &120 $\times$ 1   & 0.2 & 8479 &0.004&6.78 &42.24&0.021&0.059 \\
NGC 4151&	2000 May 28	                &720 $\times$ 3   & 0.1 & 8136 &0.003&7.56 &42.09&0.002  &0.071\\
NGC 4593&	2016 Jul 21	                &288 $\times$ 1   & 0.2 & 14121&0.008&6.89 &42.87&0.070  &0.065\\
\enddata
\tablecomments{For NGC 4593, the date listed corresponds to the observation from the \citet{Cackett18} monitoring campaign for which we obtained the cleanest fringe correction, and only this observation was used for spectral fitting. The reverberation-based black hole masses and AGN continuum luminosities (given as $\log[\lambda L_\lambda]$ at 5100~\AA) are obtained from \cite{BentzKatz15}, and the normalized Eddington accretion rate is based on $L_{\rm Bol} = 9.26L_{\rm 5100,AGN}$. PID lists the HST program ID for each observation. The extinctions ($A_{\rm V}$) are based on  \citet{Schlegel98} and assume $R_{V}=3.1$.}
\end{deluxetable*}

\begin{figure}
\centering
\includegraphics[width=9.cm]{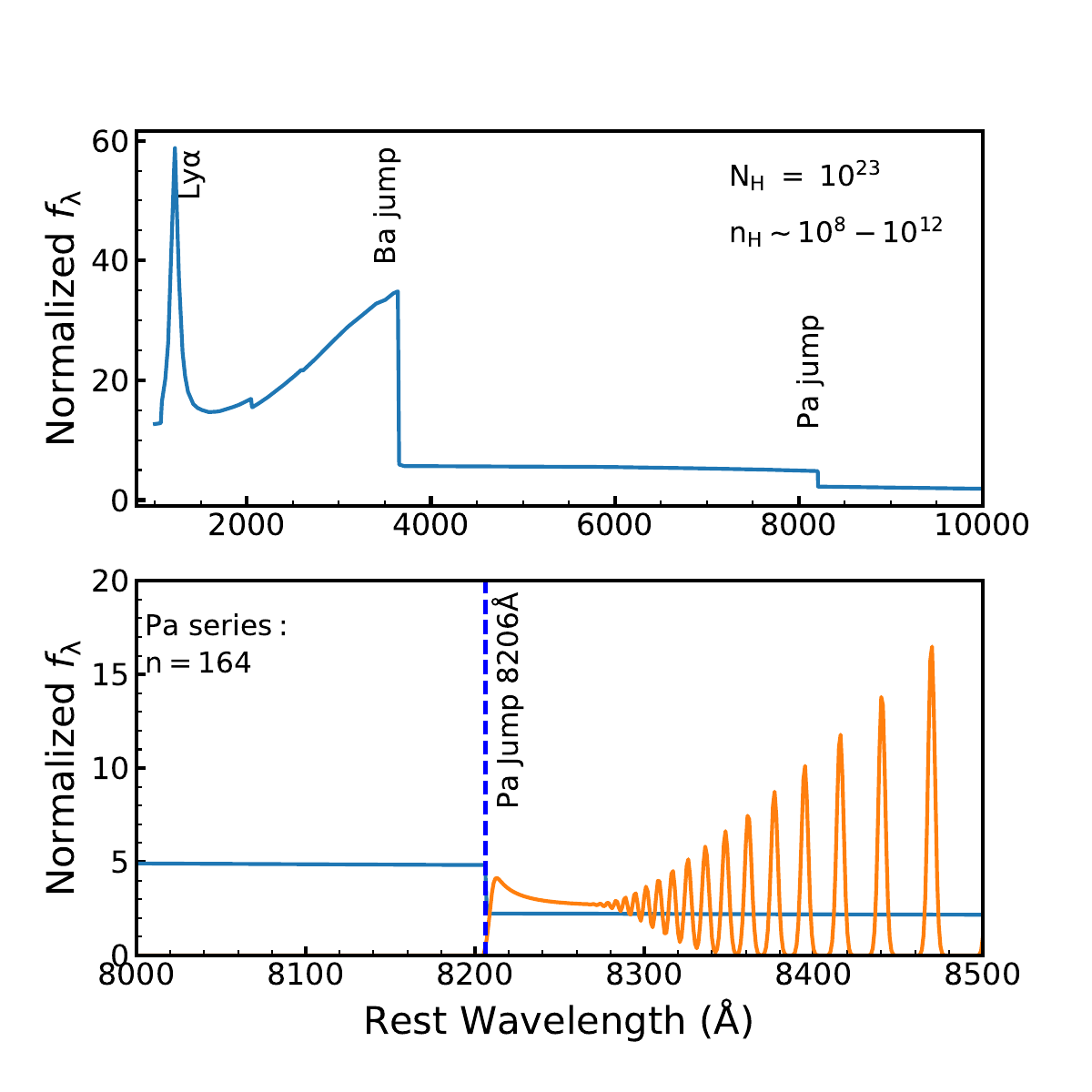}
\caption{Models of diffuse continuum (DC) emission and high-order Paschen lines. Upper panel: DC spectrum predicted by photoionization models for NGC 5548 \citep{Korista19}, used in our model fits. Lower panel: the DC model and Paschen-series emission lines up to $n = 164$, (BLR) over the range 8000--8500~\AA. The Paschen lines have been broadened with a Gaussian kernel with $\sigma = 65$ \kms.}
\label{fig:model}
\end{figure}

\subsection{Spectral decomposition method}
\label{sec:method}
We fit models to the data using the spectral fitting code {\tt PyQSOFit} \citep{Guo18,Shen19}, modified to incorporate additional model components. All of the components, including accretion disk emission (modeled as either a power law or a standard thin disk model), nebular DC, host galaxy starlight, \FeII\ emission, and high-order Paschen lines with blended \SIII\ lines, are modeled together from 6800 to 9700~\AA\ after masking out unrelated prominent emission lines, and the model is optimized by $\chi^2$ minimization.

For the DC emission, we use a model spectrum from \cite{Korista19}. This spectrum is based on local optimally-emitting cloud photoionization models for the BLR in NGC 5548, assuming a 100\% cloud coverage of the source's sky, hydrogen column density $\log(N_\mathrm{H}/\mathrm{cm}^{-2}) = 23$, and hydrogen gas density $\log(n_\mathrm{H}/\mathrm{cm}^{-3})$ integrated from 8 to 12 dex. For this work, we use only the red portion of this model ranging from 6800 to 9700~\AA. In this wavelength range, the DC spectrum primarily consists of free-bound and free-free emission. Figure \ref{fig:model} displays the full DC model over 1000--10000~\AA, including the strong scattered Ly$\alpha$ feature and Balmer jump as well as the Paschen jump. The model fitting routine modifies the DC spectrum by two free parameters: the Doppler velocity broadening and the flux normalization factor. Velocity broadening is implemented through convolution by a Gaussian kernel after rebinning the DC spectrum to a linear grid in $\log\lambda$. Following convolution, the broadened spectrum is rebinned to match the wavelength scale of the data.

High-order Paschen lines need to be included for accurate fitting of this spectral region \citep{Malkan82,Wills85,Guseva06,Reines10}. An important parameter here is the BLR gas density, because high gas densities set an upper limit to the number of Paschen series lines and also shift the wavelength of the Paschen jump slightly redward. Ionization potentials are calculated for isolated atoms, but at high densities they are subject to the electric fields of other atoms/ions. As density increases, the interaction between the electric field strengths of other atoms/ions becomes increasingly important, and the number of bound levels diminishes. The ionization potential is subsequently lowered, and this shifts the Paschen jump and the limit of the high-order lines to longer wavelengths. Here we consider Paschen lines from Pa$\epsilon$ ($n =~8$, 9548.5~\AA, where $n$ represents the upper energy level of the electron transition) to $n =$ 164 for the BLR. This upper limit of $n=164$ for bound states is an estimate corresponding to a gas density of $n_\mathrm{H} = 10^{10}\ \mathrm{cm}^{-3}$, where levels at larger $n$ would have bound state orbit size larger than the typical separation between atoms. The flux ratios of the first 50 Paschen lines are derived from \cite{Hummer87}, assuming a typical BLR with an electron temperature of 15,000 K and a density of $\rm 10^{10}\ cm^{-3}$ (case B). However, $n = 50$ may still be insufficient in some scenarios \citep{Kovacevic14} since the slope formed by the blended high-order lines starts to decrease before it reaches the Paschen edge, especially when the line dispersion is not large enough to cover the discontinuity. We further extrapolate it to 164 ($\lambda=8208.66$~\AA) by fitting a polynomial to the line intensities up to $n = 50$. The Paschen-series line spectrum is shown in the lower panel of Figure \ref{fig:model}. In the model fits, the Paschen emission-line spectrum shares the same Doppler velocity broadening as the DC spectrum, while its overall flux normalization is a free parameter. 

The high density gas will also shift the Paschen continuum jump slightly redward: shifts of  $+3.1$ \AA\ for the free-bound limit and of  $+0.31$ \AA\ for the high-order Paschen lines are expected. To simplify the spectral modeling, considering that these the small wavelength shifts make almost no difference in the fitting results for spectral models broadened to match the velocity broadening of the BLR (several thousand km s$^{-1}$), we neglected these wavelength shifts in the following analysis.

Other continuum components include the accretion disk spectrum, host-galaxy starlight, and dust emission. The accretion disk spectrum is modeled using either a power-law (PL) spectrum (a reasonable first approximation over this limited wavelength range) or a standard thermally emitting accretion disk model \citep[][hereafter SSD]{Shakura73} with an outer radius ranging from 500 to 10,000 Schwarzschild radii ($R_{\rm g}$, allowed to vary as a free parameter) and a fixed inner radius of 3$R_{\rm g}$. The power-law slope of the $F_\lambda$ spectrum from the SSD model in our fitting range is around $-7/3$, slightly depending on the black hole mass ($M_{\rm BH}$) and the normalized Eddington accretion rate ($\dot{m}$), which is estimated from the AGN continuum luminosity at 5100~\AA\ by assuming $L_{\rm Bol} = 9.26L_{\rm 5100, AGN}$ \citep{Richards06}. The reverberation-based BH mass, continuum luminosity at 5100~\AA, and $\dot{m}$ of each object are listed in Table \ref{tab:sample}. The potential contribution of starlight is modeled using simple stellar population (SSP) model spectra from \citet[][hereafter BC03]{Bruzual03}, allowing for a linear combination of model spectra spanning a range in age to achieve the best fit to the data. To model emission from hot dust, which is expected to add a small amount of the continuum flux in this spectral region \citep{Honig14}, we use a single-temperature blackbody model\footnote{Although there will be continuum contributions from dust with a range of grain sizes and chemical compositions, and from thermally emitting gas \citep{Korista19}, we use the simplest model to represent the dust emission due to the limited wavelength coverage of our STIS data.}, with a free temperature in the range 1200--1900~K \citep{Netzer15} and a free normalization factor. 

Emission lines other than the Paschen-series lines were mostly masked out from the fits, except for two [\ion{S}{3}] lines blended with high-order Paschen lines. These [\ion{S}{3}] lines  ($\lambda\lambda$9068, 9531~\AA) were modeled as single Gaussians.  We do not include iron emission templates in these fits since \ion{Fe}{2} emission lines do not contribute significantly beyond 7000~\AA.  

\begin{figure*}
\centering
\hspace*{-1cm}
\includegraphics[width=20.cm]{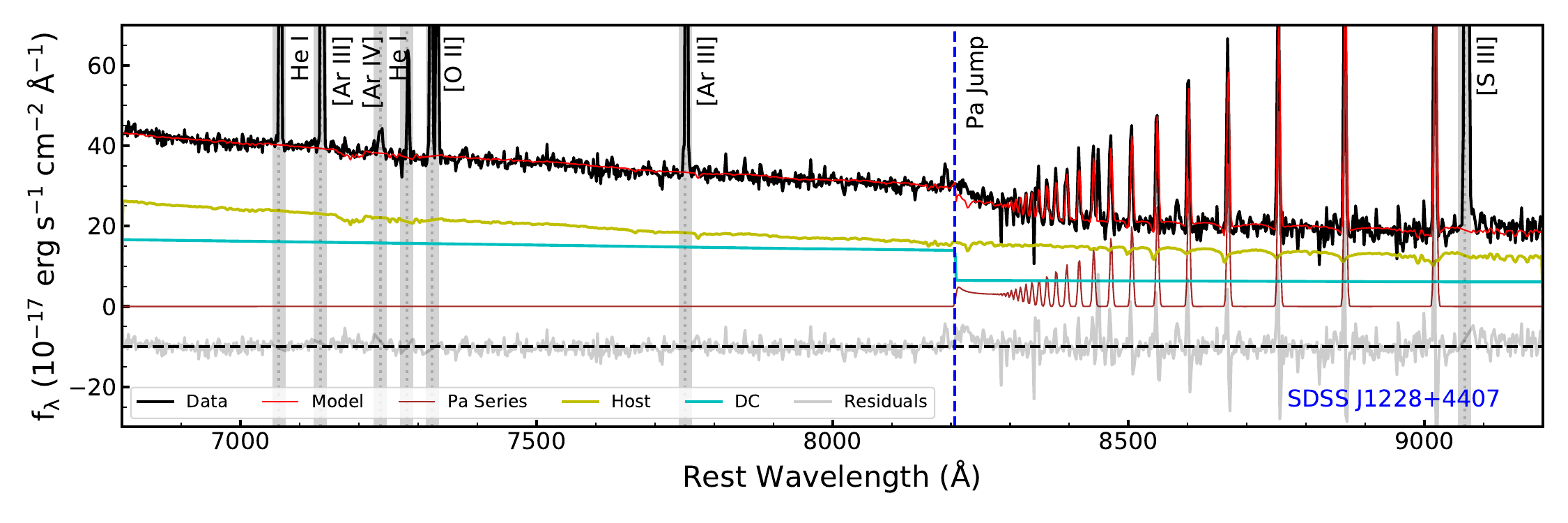}
\caption{Spectral fitting for SDSS J1228$+$4407, a star-forming galaxy at $z = 0.0007$ with a known Paschen jump from nebular continuum emission. The dereddened spectrum (black) is modeled with a synthetic host component (yellow) from BC03, diffuse continuum (DC) from Starburst99 (cyan), and Paschen lines to n = 300 (brown). The residuals (grey) are shifted downward for clarity and unrelated emission lines are masked. The blue dashed line indicates the Paschen jump at 8206~\AA.}
\label{fig:sdss}
\end{figure*}

\begin{figure*}
\centering
\includegraphics[width=17.3cm]{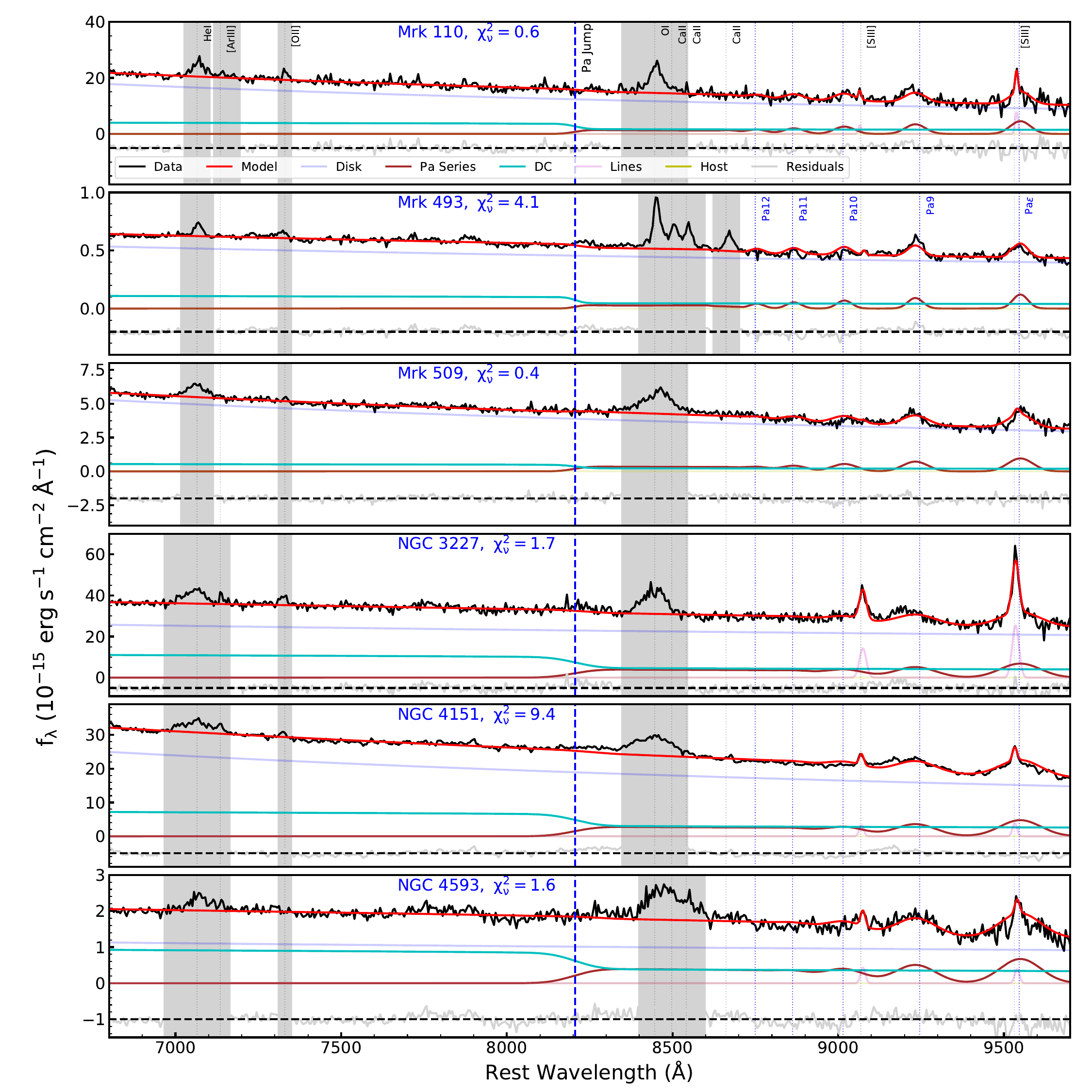}
\caption{Spectral decompositions of the six AGN in our sample, for the PL$+$SingleG model fits. The components are the same as in Figure \ref{fig:sdss}, except for a power-law component for disk and single Gaussian profiles for blended [\ion{S}{3}] lines. In these fits, the hosts are modeled with a 5 Gyr-old SSP model, but with zero flux in all objects. Primary emission lines and Paschen lines (Paschen jump) are marked with grey and blue dotted (dashed) lines, respectively. Some fringing residuals are still apparent at the longer wavelengths.
}
\label{fig:fit}
\end{figure*}

\begin{figure*}
\centering
\includegraphics[width=18.cm]{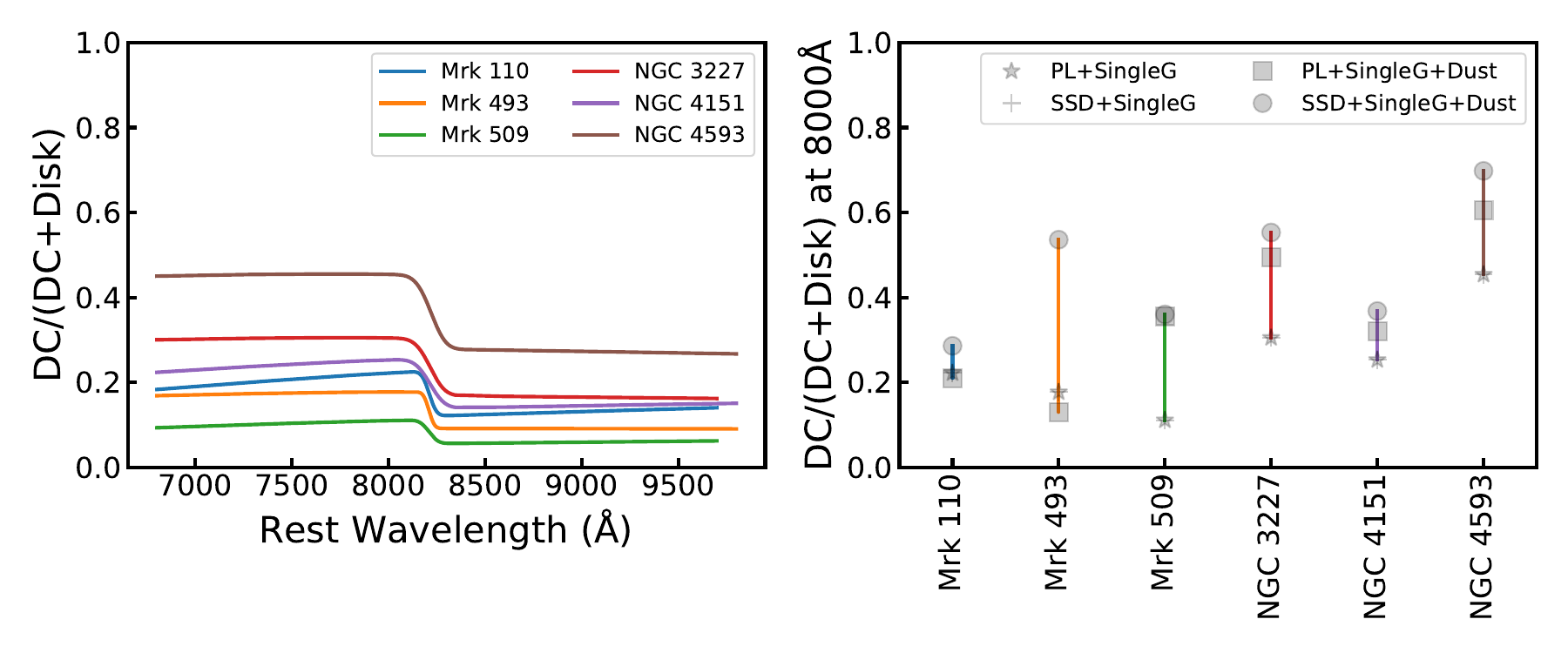}
\caption{\emph{Left:} The fractional contribution of DC component as a function of wavelength, from the PL$+$SingleG model. The DC contribution at the wavelengths shortward of the Paschen jump is around 10\% to 50\% with respect to the total DC$+$disk continuum. \emph{Right:} The range in DC fraction at 8000~\AA\ for each AGN, based on the range of model results listed in Table \ref{tab:models}.}
\label{fig:frac}
\end{figure*}

\begin{figure*}
\centering
\includegraphics[width=17.3cm]{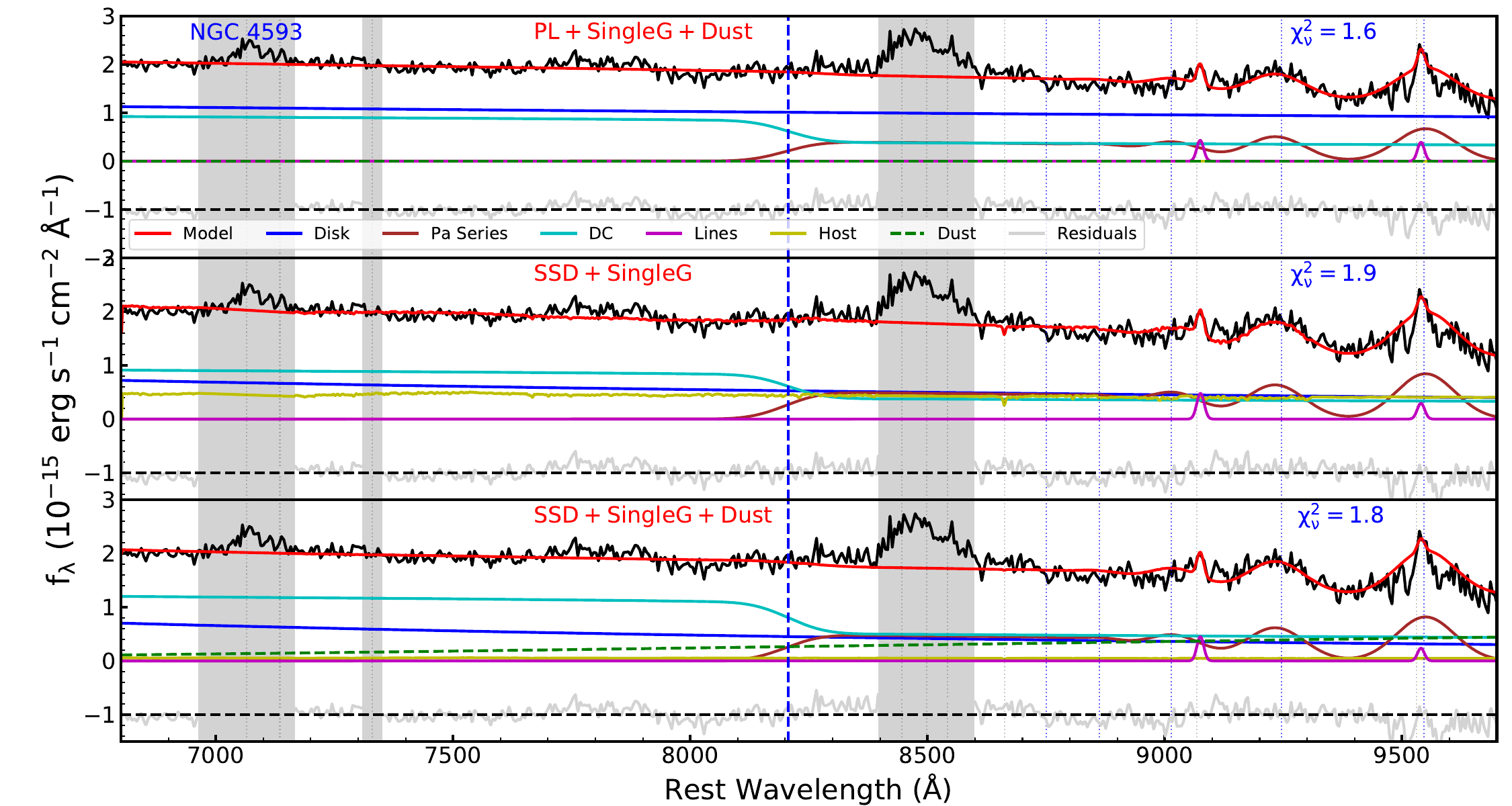}
\caption{\textbf{Comparison of spectral decompositions of NGC 4593 with different model components. The three panels illustrate the PL$+$SingleG$+$dust, SSD$+$SingleG, and SSD$+$SingleG$+$dust model fits as listed in Table \ref{tab:models}. The PL+SingleG+dust model fit is identical to the PL+SingleG fit shown in Figure \ref{fig:fit} since the model fit drives the dust component normalization to zero.  For these three models, the DC fractions with respect to DC$+$Disk fluxes are 45.4\%, 60.5\%, and 69.8\%, respectively.}}
\label{fig:varmodels}
\end{figure*}

\section{Spectral fitting results}\label{sec:results}

\subsection{Star-forming galaxy with a known Paschen jump}\label{sec:sdss}
Star-forming galaxies often exhibit a strong Paschen jump originating from \ion{H}{2} regions \citep[e.g.,][]{Guseva06,Reines10}. As a test of our model-fitting approach, we first apply the fitting method to a Galactic-extinction-corrected spectrum of a star-forming galaxy in the rest frame. For this test, we selected one galaxy with a clear Paschen jump, SDSS J1228$+$4407 at $z$ = 0.0007 (SDSS Plate-MJD-FiberID: 1371-52821-0059), from the sample of \citet{Guseva06}. To model the host-galaxy stellar component, we used a linear combination of 39 SSP models from BC03 and allowed the fit to optimize the weights for each component. As shown in Figure \ref{fig:sdss}, an obvious spectral break is present at the Paschen edge.

For this model fit, we incorporated a separate nebular continuum spectral model appropriate to the physical conditions of a star-forming galaxy (rather than an AGN BLR) using the Starburst99 models\footnote{\url{https://www.stsci.edu/science/starburst99/docs/default.htm}} \citep{Leitherer99}. The Starburst99 model is based on an instantaneous burst of $10^4~ M_{\odot}$ with a Kroupa initial mass function (0.1$-$100$M_{\odot}$) \citep{Kroupa01}, the Geneva evolutionary tracks with high mass loss, and the Pauldrach/Hillier atmospheres \citep{Hillier98,Pauldrach01}. The Paschen-series emission lines were also separately obtained from \cite{Hummer87} using case B with an estimated electron temperature of 10,000 K from \cite{Guseva06} and a density of $\rm 10^2\ cm^{-3}$ estimated from the [\ion{S}{2}] line ratio $f(6716)/f(6731) \sim  1.3$ \citep{Osterbrock89}. The high-order Paschen lines are extrapolated to $n = 300$ ($\lambda=8206.07$~\AA) rather than the $n=164$ limit used for our BLR models, due to the lower gas density in the \ion{H}{2} region environment. In the best-fitting model, the starlight component is dominated by two SSP models, having stellar population ages of 900 Myr and 9 Gyr, with a flux ratio of $\sim$2:1 at 8000~\AA, basically dominated by young stars.

Figure \ref{fig:sdss} shows that the model fit to the star-forming galaxy spectrum successfully matches the overall continuum shape across the Paschen jump region as the sum of a starlight spectrum and the nebular continuum, except in a narrow ``notch'' in the model where the opposite spectral jumps from the nebular continuum and the blended high-order Paschen lines intersect. This small gap in the model may be the result of an imperfect match of the model to the velocity broadening of the DC component, or an imperfect estimation of the blended high-order Paschen lines based on our simple extrapolation from the $n\leq50$ line intensities given that our model fit slightly underestimates the peak fluxes of the individual Paschen lines at $\lambda>8350$~\AA. Additionally, the absorption component around the Paschen jump in the synthetic host could partially account for the flux deficit. In the model fit, the DC contribution is nearly equal to the host-galaxy flux at wavelengths just below 8206~\AA, consistent with model-fitting results from other studies of other similar targets \citep{Reines10}. The strong nebular Paschen jump feature in this galaxy spectrum, easily visible in total flux, stands in strong contrast to the AGN spectra that we discuss below.

\subsection{Model fits to AGN spectra}
Figure \ref{fig:fit} displays the STIS spectra of the six AGN in our sample. Unlike the star-forming galaxy spectrum, there is no obvious sign of a jump or discontinuity at the Paschen edge in any of these spectra. To a first approximation, the continuum in each object appears smooth and featureless across this wavelength range. Here we apply the spectral decomposition method outlined above in Section \ref{sec:method} to these six STIS spectra. 

To test the sensitivity of the results to the assumptions made in constructing the model, we carried out several iterations of the fit with different model components. In each case, the DC component and high-order Paschen lines were incorporated as described in Section \ref{sec:method}.  First, a simple model consisting of a power-law (PL) AGN continuum and a single age stellar population model (``SingleG'') was tested as a reference. We expect that the STIS spectra are AGN dominated and the host contribution should be small due to the small aperture size adopted: for example, \cite{Vincentelli21} estimated that the host contribution to the Mrk~110 STIS spectrum over 7000--9500~\AA\ is less than 5\%. Furthermore, we detect no evidence of high-order Paschen absorption lines that might be expected from a post-starburst stellar population. Thus, we model the host using a single 5 Gyr-old, solar-metallicity stellar population model, since that is generally expected to be present in the host galaxy bulge and there is little spectral difference between the 5 Gyr-old and other older SSP models. Then we considered variations of this model that included (1) a more physically motivated disk component by using the SSD continuum instead of a single PL; (2) a contribution of thermal dust emission from the torus. The model combinations that we tested include: PL$+$SingleG$+$Dust, SSD$+$SingleG, and SSD$+$SingleG$+$Dust.

Focusing primarily on the continuum components in the fit, we first mask out unrelated broad and/or narrow emission lines, including \ion{He}{1} $\lambda\lambda$7065,7281, [\ion{Ar}{3}] $\lambda$7136, [\ion{Ar}{4}] $\lambda$7171, [\ion{O}{2}] $\lambda\lambda$7320,7330, \ion{O}{1} $\lambda$8446, and \ion{Ca}{2} $\lambda\lambda\lambda$8498,8542,8662, where these lines are found to be present in the STIS data (see Figure \ref{fig:fit}). The fits are then carried out over the rest wavelength range 6800 -- 9700~\AA, over a total number of pixels between 1541 and 1586 (after masking) for each object.

The spectral fitting results for models with a power-law AGN continuum and single stellar population are displayed along with the STIS data in Figure \ref{fig:fit}. As expected, the host fraction in each object is zero if considering a 5 Gyr-old SSP model. The model fits also demonstrate that, while the DC contribution can be significant shortward of the Paschen jump (also see below and Table \ref{tab:models}), its ``step-down" feature can be almost perfectly offset by the excess of the blended and broadened high-order Paschen lines. These results demonstrate that a smooth and featureless continuum can still be compatible with a substantial contribution of DC flux around 8000~\AA, as predicted by \citet{Malkan82}.

The left panel of Figure \ref{fig:frac} illustrates the fractional contribution of the DC component with respect to the total flux for these initial model fits. The DC fraction from the BLR at 8000~\AA\ ranges from $\sim$ 10 to 50\% in different objects, with a median value of $\sim$ 20\%. This significant contribution of DC is consistent with discoveries in star-forming galaxies \citep{Guseva06,Reines10}. However, the results of these fits are not unique, as the inferred DC fraction depends on the assumptions made for the other continuum components (AGN disk emission, starlight, and dust). Table \ref{tab:models}  and the right panel of Figure \ref{fig:frac} illustrate the range in the inferred DC fraction for each object for the different model variants described above. The range in fitting results for different model variants indicates that the uncertainty in the DC fraction is dominated by the choices made for other continuum components, similar to the conclusions of \citet{Vincentelli21}. This range is a factor of $\sim$ 1 to 3 for the six objects, much larger than the typical statistical uncertainties in DC fraction ($\lesssim$ 5\%) from the Monte Carlo error analysis procedure applied to each model fit. It is worth noting that the 22 -- 29\% DC fraction of Mrk 110 is consistent with the 12 -- 30\% DC fraction in $i$ band found by \cite{Vincentelli21}, indicating that our Paschen jump fit yields a similar result as the full spectral fit over 1500 -- 10000~\AA\ in this object. (The Vincentelli et al.\ model fit to the STIS spectrum of Mrk 110 did not include blended high-order Balmer and Paschen lines; instead, the regions around the Balmer and Paschen jumps were simply masked out from their fit.)

The choice of disk spectrum (PL or SSD) substantially alters the DC fraction in all cases, while the dust contribution is usually minor (almost zero for PL) and only slightly affects the DC fraction, except in Mrk 493 for the SSD model fit. The best-fitting PL slopes (close to the observed spectral slope, see Table \ref{tab:models}) in all cases are much shallower than the expected slope of $-7/3$ from the SSD model such that the spectral fitting can only obtain a satisfactory fit with the SSD continuum by lowering its contribution. Only in Mrk 110 is the DC fraction similar between the PL and SSD models, primarily due to the relatively steep spectral slope of the data.  In addition, if we employ more complex SSP models\footnote{We also tested more complex SSP models: a linear combination of three SSP models with different ages of 100 Myr, 900 Myr, 5 Gyr, together with the PL model. The young stellar population only has small contributions in NGC~3227 and NGC~ 4151, which increases the DC contribution by a fraction of $\sim$ 3\% to 6\%.} that include a young stellar population that could mimic the AGN disk continuum, this will naturally reduce the disk fraction and therefore increase the relative DC contribution. We here consider the PL$+$Single model as our fiducial result as this model is the simplest, and it yields the best reduced $\chi^2$ for all six objects. Moreover, it is worth emphasizing that we can consider the DC fractional contribution in PL$+$SingleG model as a lower limit based on the current DC model from \citet{Korista19}, since it gives the necessary fraction to balance the high-order Paschen lines in the simplest model if no obvious Paschen jump appears in total flux, and any additional components will usually reduce the disk contribution and thus increase the DC fraction.  

To illustrate an example of the differences in fits for the various choices of model components and the resulting differences in DC fraction, Figure \ref{fig:varmodels} shows three fits (PL+SingleG+dust, SSD+SingleG, and SSD+SingleG+dust) to the NGC 4593 spectrum. The first model (PL$+$SingleG$+$Dust) is identical to the PL$+$SingleG model in Figure \ref{fig:fit} since the dust component normalization is zero in the PL+SingleG+Dust fit. When the SSD model is employed, the fits result in stronger host and dust components, reducing the disk flux and hence increasing the DC fraction.


One important note of caution for interpretation of these fitting results is that the DC model is based on photoionization modeling tailored to the properties of NGC 5548 \citep{Korista19} and not computed specifically for each object in our sample. Differences in physical conditions within the BLR, such as the gas metallicity, and distributions of gas number density and column density, can alter the DC shape and strength \citep{Korista19}. Furthermore, the uncertainty is relatively larger in estimating the ionizing photon flux based on an optical continuum measurement, which will be used in constructing the photoionization model to map the ionized flux to radius in flux density plane. Our modeling results are dependent on the simplifying assumption that the DC spectral shape and hydrogen ionizing photon flux are intrinsically similar among the objects in our sample, other than its fractional contribution to flux and its velocity broadening. Moreover, a single PL fit is based on the asymptotic spectral energy distribution (SED) behavior of a disk with infinite radius, and is not expected to be an accurate model for the actual disk continuum except as an approximation over very restricted wavelength ranges. A real disk SED is expected to turn downward at wavelengths in the near-IR, asymptoting toward the Rayleigh-Jeans tail of the last annulus in the disk. This would allow for potentially greater contributions from the DC and/or the dusty torus. Finally, winds originating from the central accretion disk region and BLR would produce faint broad wings to lines but would be otherwise hard to detect \citep{Dehghanian20}. The optical/NIR emission from the wind could be significant and would manifest as warm free-free emission with recombination features superimposed.



\subsection{Validation of DC models}
\subsubsection{Comparing with photoionization predictions}
One important consistency check for these spectral fitting results is whether the inferred DC strength is compatible with the fluxes of BLR emission lines such as Ly$\alpha$ and the Balmer lines. We can carry out this check using photoionization modeling results from \citet{Korista19}.

To complement the STIS G750L spectra for our AGN sample, we also collected and combined the available STIS data in the G140L and G430L grating settings, for observations taken contemporaneously with the G750L data. After correcting for the Galactic extinction, we conducted local fits to the spectral regions surrounding Ly$\alpha$ and \hb\ with {\tt PyQSOFit} following methods similar to \cite{Shen19}. The broad Ly$\alpha$ and \hb\ are separated from the narrow components for luminosity estimation. Table \ref{tab:line} presents the measurements of these line luminosities along with continuum luminosities at 1215~\AA\ ($L_{\rm 1215}$). The Ly$\alpha$ line for Mrk 110 is unavailable due to the lack of G140L spectra. The continuum luminosity of Mrk 110 at 1215~\AA\ is estimated by extrapolating the power-law continuum from a near-UV spectral fit to the STIS G230L data. The far-UV spectrum of NGC 3227 (both continuum at 1215~\AA\ and Ly$\alpha$) is heavily reddened by small dust grains intrinsic to the AGN \citep{Crenshaw01,Mehdipour21}, especially at wavelengths below 2000~\AA, and we thus exclude it in our comparison with photoionization models. The blue wing of Ly$\alpha$ is strongly affected by broad and/or narrow absorption lines for all objects in our sample, and its red wing is blended with \ion{N}{5} $\lambda$1240. Despite masking these absorption lines, it is still very challenging to estimate precisely the intrinsic line profile and to separate the broad and narrow emission components. Consequently, the measurement uncertainty of Ly$\alpha$ is estimated to be $\sim$ 20\%, much higher than those of \hb\ and $L_{\rm 1215}$ ($\lesssim$ 5\%).

To compare the DC fluxes with H$\beta$, we extrapolated the DC models (see Table \ref{tab:models}) for each object based on the Paschen jump region spectral fits to derive the DC luminosities at 5200~\AA. The DC luminosity ranges (based on the set of model fits listed in Table \ref{tab:models}) are listed in Table \ref{tab:line}), where $L_\mathrm{DC,5200}$ refers to $\lambda L_\lambda$ for the DC component at $\lambda=5200$~\AA.

\begin{figure}
\hspace{-1cm}
\centering
\includegraphics[width=9.cm]{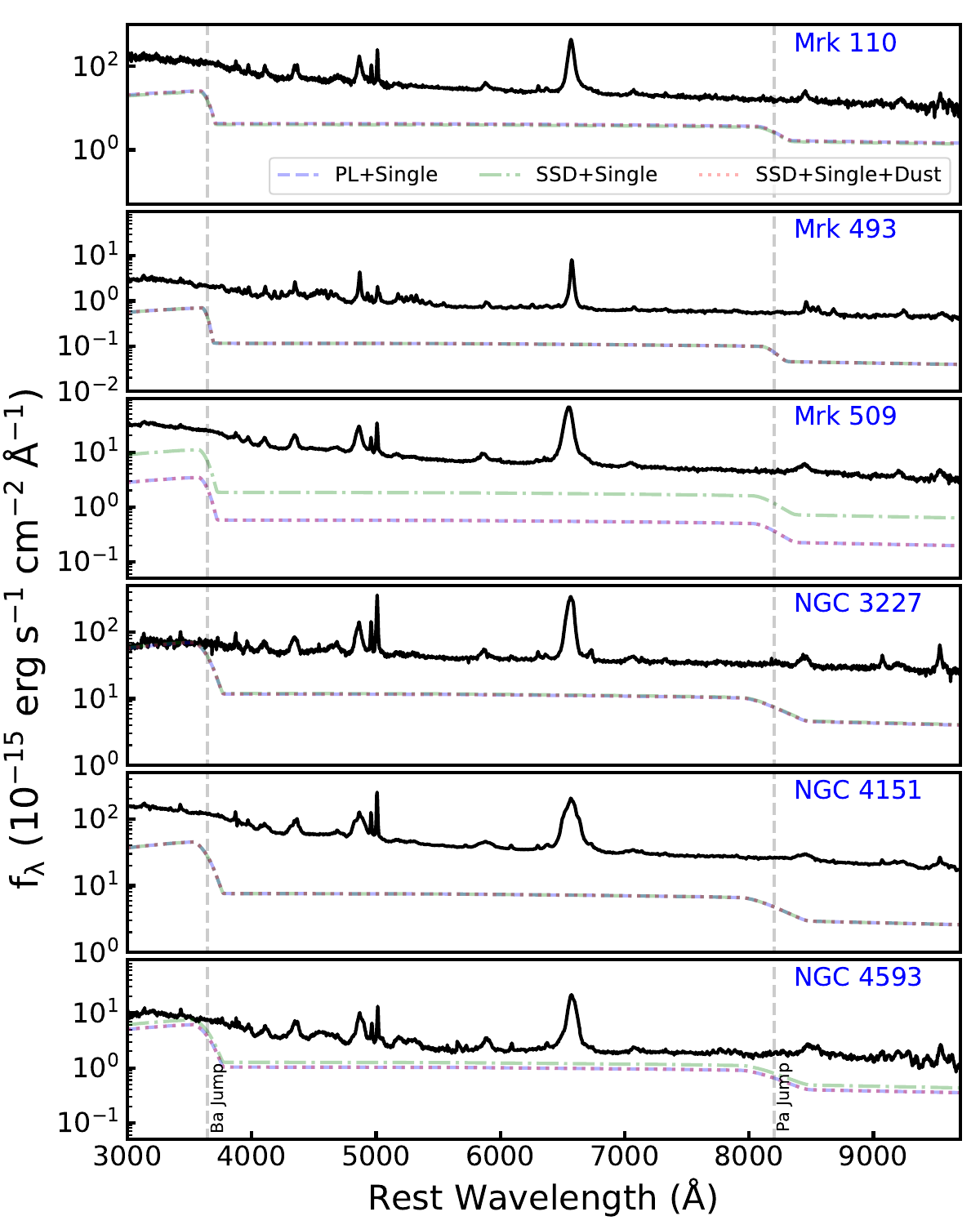}
\caption{STIS spectra of the six AGN in our sample, shown with extrapolated DC fluxes from different models. The STIS spectra are obtained from the G430L and G750L gratings. The DC spectra correspond to different model fits listed in Table \ref{tab:models}. The DC components are velocity broadened according to their original fitting results. The Balmer and Paschen jumps are marked with dashed grey lines.}
\label{fig:check}
\end{figure}

For NGC 5548, \cite{Korista19} predicted that the ratio $L_{\rm DC, 5200}/L_{\rm incident, 1215}$ is $\sim$ 0.2 in a typical BLR environment (see their Figure 1, 9 and our Table \ref{tab:line}). Furthermore, the DC luminosity at 5200~\AA\ should be comparable to the line luminosity of Ly$\alpha$ (e.g., $L_{\rm DC, 5200}/L_{\mathrm{Ly}\alpha} \sim $ 1) and the line flux of \hb\ is expected to be one order of magnitude lower than that of Ly$\alpha$ or $L_{\rm DC, 5200}$ \cite[see Figure 2 in][]{Korista19}. These ratios are approximate and also depend on the BLR environment assumptions (e.g., gas density, gas distribution, and metallicity), but they provide a general consistency check on the DC strengths derived from our model fits to the STIS data. In our comparison, we assume the total continuum luminosity at 1215~\AA\ is similar to the incident one (e.g., $L_{\rm 1215} \approx L_{\rm incident,1215}$), because \citet{Korista19} predict that the diffuse/total continuum flux ratio is $<0.1$ in the continuum region around the Ly$\alpha$ line.

In Table \ref{tab:line}, we list the measured luminosity ratios of our sample, as well as those of NGC 5548 as a reference. Aside from the anomalous case of NGC 3227 which is impacted by heavy extinction, the ratios are $L_{\rm DC, 5200}/L_{\rm 1215}$ $\sim$ 0.1 $-$ 0.6, $L_{\rm DC, 5200}/L_{\mathrm{Ly}\alpha}$ $\sim$ 1 $-$ 5, and $L_{\rm H\beta}/L_{\rm DC, 5200}$ $\sim$ 0.1 $-$ 0.3, respectively. These are broadly consistent with the photoionization predictions considering the potential AGN host reddening, different photoionization properties, and measurement uncertainties of Ly$\alpha$. This general consistency between the predicted and observed luminosity ratios between the DC strength and BLR hydrogen lines provides additional support for our conclusions regarding the range of inferred DC strengths in the Paschen jump region.

\subsubsection{Comparing with total UV/optical flux}
Another important consistency check is to use the results of our model fits to extrapolate the inferred DC flux to shorter wavelengths, and compare with the observed AGN flux. If the DC flux greatly exceeds the total observed flux at shorter wavelengths (e.g., below the Balmer jump), this would indicate that our method of fitting models to the Paschen jump region does not provide an accurate indication of the true DC flux level. If the extrapolated DC component accounts for a substantial fraction of the observed continuum flux below the Balmer jump, without exceeding the observed flux, then the normalization of the DC component inferred from the Paschen-jump region may be reasonable.

Figure \ref{fig:check} presents a broader spectral range for our sample of AGN from STIS data in the  G430L and G750L gratings, together with the extrapolated DC flux from three versions of our model fits.  (The DC fluxes are identical in the PL$+$SingleG models with or without dust included, so we present only the dust-free version of the model fit). For the three models presented, the DC fluxes are usually very similar since adding additional components only changes the relative contribution of the disk (PL or SSD), except in the case of Mrk 509 where the normalization of the DC component in the SSD+Single model differs substantially from the other two model fits. The DC component reaches its maximum in $f_\lambda$ around the Balmer edge, then decreases towards shorter wavelengths as shown in Figure \ref{fig:model}. We found that the DC fluxes are all comparable to or significantly lower than then total flux around the Balmer edge, indicating that our model results pass the qualitative consistency check described above. For the case of NGC 3227, the extrapolated DC flux just below the Balmer jump is essentially equal to the total observed continuum flux. However, NGC 3227 has substantial internal extinction within the AGN host that has not been corrected in the STIS data, and this extinction suppresses the observed flux at blue wavelengths. 

In principle, carrying out multicomponent fits to the full STIS wavelength range would provide more stringent constraints on the DC contribution in these AGN. However, complete broad-band spectral fitting from the UV through near-IR remains a major challenge, requiring accurate modeling of the overall disk emission spectrum including possible departures from SSD model predictions, the pile-up of high-order Balmer lines near the Balmer jump, the \ion{Fe}{2} emission blends, and internal extinction within the AGN. Such modeling is beyond the scope of this work, but the STIS data for these six objects represents the best available testing ground for broad-band AGN spectral fitting spanning Ly$\alpha$ through the Paschen jump region.


\begin{deluxetable*}{cccccc}[htb]
\tablecaption{DC fraction (\%) at 8000 \AA\ in different models \label{tab:models}}
\tablecolumns{6}
\tablewidth{0pt}
\tablehead{
\colhead{Name} &
\colhead{PL + SingleG (slope)} &
\colhead{PL+ SingleG + Dust (slope)} &
\colhead{SSD + SingleG } &
\colhead{SSD + SingleG + Dust }
}
\startdata
Mrk 110 &22.2 ($-$1.95)&22.2 ($-$1.95) &21.0 &28.6 \\
Mrk 493 &17.7 ($-$0.85)&17.7 ($-$0.85) &13.0 &53.6\\
Mrk 509 &11.1 ($-$1.62)&11.1 ($-$1.62) &36.0 &35.6\\
NGC 3227&30.5 ($-$0.61)&30.5 ($-$0.61) &49.4 &55.3 \\
NGC 4151&25.3 ($-$1.47)&25.3 ($-$1.47) &32.0 &36.8\\
NGC 4593&45.4 ($-$0.58)&45.4 ($-$0.58) &60.5 &69.8 \\   
\enddata
\tablecomments{The spectra are modeled with a power-law (PL, the slopes $\alpha$ in $f_{\lambda} \propto \lambda^{a} $ are listed in parentheses) or standard accretion disk model (SSD) for the disk component. Starlight is modeled by a single 5 Gyr-old, solar-metallicity stellar population model (SingleG), and the last two models include thermal dust emission.}
\end{deluxetable*}

\begin{deluxetable*}{cccccccccccc}[htb]
\tablecaption{Continuum and line luminosities \label{tab:line}}
\tablecolumns{12}
\tablewidth{0pt}
\tablehead{
\colhead{Name} &
\colhead{Log $L_{\rm 1215}$} &
\colhead{Log $L_{\rm Ly\alpha}$} &
\colhead{Log $L_{\rm \hb}$}&
\colhead{Log $L_{\rm DC, 5200}$}&
\colhead{$L_{\rm DC, 5200}/L_{\rm 1215}$}&
\colhead{$L_{\rm DC, 5200}/L_{\mathrm{Ly}\alpha}$}&
\colhead{$L_{\rm H\beta}/L_{\rm DC, 5200}$}\\
\colhead{} &
\colhead{(erg s$^{-1}$)} &
\colhead{(erg s$^{-1}$)} &
\colhead{(erg s$^{-1}$)} &
\colhead{(erg s$^{-1}$)} &
\colhead{} &
\colhead{}
}
\startdata
NGC 5548 & 43.6 & 42.6 & 41.4  & 42.9           & 0.20             & 2.00                & 0.32          \\
\hline
Mrk 110  & 44.1 & --   & 42.4  & 43.8 $-$ 43.9  & 0.5 $-$ 0.63      & --                  & 0.03 $-$ 0.04 \\
Mrk 493  & 43.1 & 42.3 & 40.9  & 42.1 $-$ 42.6  & 0.10 $-$ 0.32     & 0.79 $-$ 1.99       & 0.02 $-$ 0.06 \\
Mrk 509  & 44.4 & 43.5 & 42.5  & 42.9 $-$ 43.5  & 0.03 $-$ 0.13     & 0.25 $-$ 1.00       & 0.10 $-$ 0.40 \\
NGC 4151 & 43.0 & 41.8 & 41.2  & 41.9 $-$ 42.1  & 0.08 $-$ 0.13     & 1.26 $-$ 1.99       & 0.13 $-$ 0.20 \\
NGC 4593 & 42.3 & 41.4 & 40.9  & 41.9 $-$ 42.1  & 0.40 $-$ 0.63     & 3.16 $-$ 5.01       & 0.06 $-$ 0.10 \\
\enddata
\tablecomments{The emission-line luminosities are measured from STIS spectra, and the DC luminosity ranges at 5200~\AA\ are obtained from extrapolation of different DC models in Figure \ref{fig:fit}. The continuum luminosity of Mrk 110 at 1215~\AA\ is extrapolated from its PL continuum fitting of the near-UV spectrum. NGC 3227 is excluded from this comparison due to its high UV extinction. }
\end{deluxetable*}

\section{Discussion}\label{sec:dis}

The Paschen jump is one of a few prominent features of the diffuse nebular continuum spectrum expected in AGN, yet it is not apparent at all in high-quality, host-free HST spectra. None of the six AGN presented in this work exhibit any evidence of a spectral break or discontinuity at the Paschen edge at 8206~\AA, and to the best of our knowledge, no previous studies have reported the discovery of Paschen jumps in other broad-line AGN. The results of our model fits show that a plausible explanation for the smoothness of the spectra across the jump is that the DC discontinuity is smoothed out by the pile-up of high-order Paschen lines, as previously anticipated by \citet{Malkan82}, and also by the potential presence of gas at very high density \citep[see below;][]{Vincentelli21}. In comparison with star-forming galaxies exhibiting a clear Paschen jump \citep[e.g.,][]{Guseva06,Reines10,Gunawardhana20}, the Doppler velocity broadening in broad-line AGN is also a key factor in smoothing out the jump. Our fits assume the same velocity broadening for the DC and the Paschen lines as both of them are expected to raise from the BLR, and the satisfactory fitting results across the Paschen jump are consistent with the DC and the Paschen lines originating from similar radii in the BLR. However, we note that the assumption of a uniform velocity broadening for all of the Paschen-series lines is an oversimplification. Based on a model for NGC 5548, the line formation radius is expected to be smaller for higher-order lines, decreasing by a factor of $\sim2$ from the low-order to high-order lines; thus the velocity width of high-order lines should be correspondingly broader (by a factor of $\sim2^{1/2}$ under the assumption of virial motion of BLR clouds). This effect will further smooth out the Paschen edge feature contributed by the high-order lines. 

Compared with the Paschen jump, the Balmer jump excess is much more prominent in total flux spectra of AGN, but is also blended with \ion{Fe}{2} emission, high-order Balmer lines, and other emission lines. A notable feature of the small blue bump excess is that it typically extends to somewhat longer wavelengths than the 3647~\AA\ wavelength of the Balmer jump (e.g., $\sim$ 3800~\AA), as can be seen in the composite quasar spectrum of \citet{VandenBerk01}, for example. This extended ``bump" feature extending redward of 3647~\AA\ is unlikely to be due to the velocity broadening of the DC Balmer jump excess, as it would require a much larger broadening width than the widths of the Balmer emission lines. As discussed by \cite{Vincentelli21}, the extension of the small blue bump redward of the Balmer jump is likely due to the presence of high-density gas (e.g., $\rm 10^{12-13}\ cm^{-3}$) within the BLR making the free-bound jump slightly redshifted, as a result of the finite gas density effect\footnote{As previously described, the wavelength of the Paschen jump is dependent on gas density due to the reduction in the number of bound levels of hydrogen atoms in high-density gas. If there exists a broad range in gas densities in the BLR, the Paschen jump effectively becomes a series of jumps each with a different wavelength that when summed together appear as a smooth decline toward longer wavelengths, rather than an abrupt jump.} as mentioned in \S \ref{sec:method}, and the \citet{Korista19} models do not incorporate this effect. Nevertheless, our model fits show that the Paschen jump region does not appear to have a corresponding excess, and the Paschen region can be modeled adequately with the combination of a DC model and high-order Paschen lines all broadened by the same velocity width.

Recently, intensive disk RM experiments have revealed strong evidence of a $U$-band excess around the Balmer jump in the lag spectrum, as well as a mild 8000~\AA\ excess in the vicinity of the Paschen jump in some objects \citep[e.g., NGC 5548, NGC 4593, Fairall 9;][]{Fausnaugh16,Cackett18,HernandezSantisteban20}. These two excesses are important clues to the unique asymmetric Balmer and Paschen jumps of the DC component. Thus, searching for a change in reverberation lag across the Paschen jump, combined with modeling the spectral shape across the jump in both total flux and variable (rms) flux, could be a useful diagnostic to constrain the underlying accretion disk lag spectrum and test disk reprocessing models \citep{Korista19}. However, a complete physical interpretation of the lag spectra would require models that account for the lags of all of the variable components in this region, including the disk, the DC, the high-order Paschen lines, and the dust continuum. Our results indicate a lower limit to the DC fraction, based on the specific set of spectral models considered here, ranging from 10\% to 50\% with respect to the total continuum flux, and this DC contribution should appreciably affect the continuum lag measurements (particularly if it is at the upper end of this estimated range), to a degree that depends on physical properties of the BLR gas as well as on the variability amplitude and characteristic variability timescale of the driving continuum \citep{Korista19}. According to the photoionization calculations of \citet{Korista19}, in the $i$-band spectral region, the relative importance of the DC component to the measured inter-band continuum delays is nearly proportional to its fractional contribution to the total flux (see their Figure 11). This implies that the ratio of total continuum lag to disk continuum lag will be $\tau_{\rm disk+DC}/\tau_{\rm disk}$ $\sim$ 1.1 to 1.5. However, quantitative estimation of $\tau_{\rm disk}$ will require both specific photoionization modeling that is beyond the scope of this paper and variability models incorporating the lag response of all time-varying spectral components,  and this work will be deferred to a future analysis incorporating broader spectral coverage spanning the Balmer and Paschen jumps. According to the recent modeling of Fairall 9 \citep{HernandezSantisteban20} and NGC 5548 \citep{Lawther18}, the DC component indeed increases the inter-band lags to some degree, but is probably still insufficient to explain the factor of $\sim3$ discrepancy between disk sizes inferred from continuum reverberation mapping and simple model predictions for standard disks \citep{Shakura73}. We note that even in the context of other disk reprocessing scenarios \citep[e.g.,][]{Kammoun21}, the contribution of the time-variable DC emission to the overall continuum lag spectrum must be taken into account for detailed comparison of models with reverberation data. Alternatively, other scenarios have been proposed in which the observed lags are dominated by the BLR rather than by the disk: \citet{Netzer21} recently examined the continuum lag spectra of several AGN and concluded that reprocessing by radiation pressure confined BLR clouds can entirely explain the observed lag-spectra both in shape and magnitude, without requiring any lag contribution from the disk. In another work, \citet{Gaskell17} suggested that internal extinction within the AGN, which can be traced by the broad-line Balmer decrement, may account for the unexpected larger disk size. They argued that the observed disk sizes of 14 local AGN from the sample of \citet{Sergeev05} could be comparable with those predicted by AGN continuum luminosities (see their Figure 2) if a correction for extinction is carried out. According to their calculations, for these 14 AGN with \ha/\hb\ $\sim$ 4, the intrinsic flux at 5100 \AA\ could be $\sim$ 1 mag brighter by assuming the unreddened Balmer decrement to be 2.72 and a Milky Way extinction curve. This suggests that our sample may also be slightly affected by internal AGN extinction, with an average broad \ha/\hb\ $\sim$ 3.5 for five objects (excluding NGC~3227 with \ha/\hb $\sim$ 4) obtained from the full spectral fitting.

\section{Conclusions}\label{sec:con}
We performed spectral decompositions for six nearby type 1 AGN at $z$ $<$0.1 to evaluate the fractional contribution of the nebular diffuse continuum \citep[DC,][]{Korista19} to the total flux. HST STIS spectra taken with narrow slit apertures exclude the vast majority of the circumnuclear starlight from the data, allowing a much cleaner test for the presence of a Paschen jump than would be possible with ground-based data. Our main findings are as follows:
\begin{enumerate}
    \item In each case, the Paschen jump is imperceptible in the spectra of these unobscured AGN. Our model fits imply that the DC Paschen jump, which is expected to be present in the data, can be balanced by the excess flux of high-order Paschen lines in the vicinity of 8206~\AA\ (Figure \ref{fig:fit}), as expected by \cite{Malkan82}. These two spectral components from the BLR can combine to produce a virtually featureless and smooth total continuum across the Paschen edge. This behavior stands in contrast to the Balmer jump region, where the much stronger DC Balmer jump combined with \ion{Fe}{2} emission gives rise to the easily recognised small blue bump excess in total flux.
    
    \item The DC emission originating from the BLR makes a significant contribution to the total flux at 8000~\AA\ in our sample, at least 10\% to 50\% in different cases (Figure \ref{fig:fit} and \ref{fig:frac}), although our modeling is subject to substantial systematic uncertainty primarily due to the spectral shape of the AGN disk continuum emission. The DC emission may still be responsible for a discontinuity in continuum reverberation lag across the Paschen jump that could potentially serve as a useful diagnostic if it can be measured accurately in future reverberation mapping campaigns.
\end{enumerate}
In the future, we will further explore broad-band spectral fitting incorporating both the Balmer and Paschen jumps to further constrain the DC contribution. Furthermore, the host galaxy and torus contributions can be better constrained using data extending to longer wavelengths in the near-infrared to reduce the degeneracies in the spectral decompositions, although this could introduce other modeling challenges due to differing spatial aperture sizes for UV/optical and infrared spectroscopic data. Together, continuum reverberation mapping and detailed spectral decompositions across a broad wavelength range can provide important new insights into the physics of DC emission as well as AGN accretion disk reprocessing.

\acknowledgements
We thank anonymous referee for the helpful suggestions. We thank Matthew Malkan, Ari Laor for the useful discussions about the physics of the Paschen jump.  We acknowledge the contributions of additional co-investigators to the proposal for HST program 15124 including Rick Edelson, Michael Fausnaugh, Jelle Kaastra, and Bradley Peterson. Research at UCI has been supported in part by NSF grant AST-1907290. Support at UCI for HST programs 14744 and 15124 was provided by NASA through grants from the Space Telescope Science Institute, which is operated by the Association of Universities for Research in Astronomy, Inc., under NASA contract NAS5-26555. EC acknowledges funding support from HST program number 15413 (for Mrk 110) and NSF grant AST-1909199. LCH is supported by the National Science Foundation of China (11721303, 11991052) and the National Key R\&D Program of China (2016YFA0400702). MCB gratefully acknowledges support from the NSF through grant AST-2009230. VU acknowledges funding support from the NASA Astrophysics Data Analysis Program Grant \# 80NSSC20K0450. MV gratefully acknowledges support from the Independent Research Fund Denmark via grant number DFF 8021-00130. GJF acknowledges support by NSF (1816537, 1910687), NASA (ATP 17-ATP17-0141, 19-ATP19-0188), and STScI (HST-AR-15018 and HST-GO-16196.003-A). JMG gratefully acknowledges support from NASA under the awards 80NSSC17K0126 and 80NSSC19K1638. WNB acknowledges support from NSF grant AST-2106990.

Funding for the Sloan Digital Sky 
Survey IV has been provided by the 
Alfred P. Sloan Foundation, the U.S. 
Department of Energy Office of 
Science, and the Participating 
Institutions. 

SDSS-IV acknowledges support and 
resources from the Center for High 
Performance Computing  at the 
University of Utah. The SDSS 
website is www.sdss.org.

SDSS-IV is managed by the 
Astrophysical Research Consortium 
for the Participating Institutions 
of the SDSS Collaboration including 
the Brazilian Participation Group, 
the Carnegie Institution for Science, 
Carnegie Mellon University, Center for 
Astrophysics | Harvard \& 
Smithsonian, the Chilean Participation 
Group, the French Participation Group, 
Instituto de Astrof\'isica de 
Canarias, The Johns Hopkins 
University, Kavli Institute for the 
Physics and Mathematics of the 
Universe (IPMU) / University of 
Tokyo, the Korean Participation Group, 
Lawrence Berkeley National Laboratory, 
Leibniz Institut f\"ur Astrophysik 
Potsdam (AIP),  Max-Planck-Institut 
f\"ur Astronomie (MPIA Heidelberg), 
Max-Planck-Institut f\"ur 
Astrophysik (MPA Garching), 
Max-Planck-Institut f\"ur 
Extraterrestrische Physik (MPE), 
National Astronomical Observatories of 
China, New Mexico State University, 
New York University, University of 
Notre Dame, Observat\'ario 
Nacional / MCTI, The Ohio State 
University, Pennsylvania State 
University, Shanghai 
Astronomical Observatory, United 
Kingdom Participation Group, 
Universidad Nacional Aut\'onoma 
de M\'exico, University of Arizona, 
University of Colorado Boulder, 
University of Oxford, University of 
Portsmouth, University of Utah, 
University of Virginia, University 
of Washington, University of 
Wisconsin, Vanderbilt University, 
and Yale University.

\bibliography{ref.bib}
\bibliographystyle{aasjournal}

\end{CJK}
\end{document}